 \documentclass[final,1p,times,twocolumn]{elsarticle}
\usepackage{dcolumn}%
\usepackage{bm}%
\usepackage{graphicx}
\usepackage{amsfonts}
\usepackage{amsmath}
\usepackage{amssymb}
\usepackage{hyperref}
\usepackage{indentfirst}
\usepackage[nodots]{numcompress}
\usepackage[usenames,dvipsnames]{pstricks}
\usepackage{epsfig}
\usepackage{pst-grad} 
\usepackage{pst-plot} 
\journal{Physics Letters A}

\begin{document}
\begin{frontmatter}

\title{Coupling between magnetic field and curvature in Heisenberg spins on
surfaces with rotational symmetry}


\author[1]{Vagson L. Carvalho-Santos}
\address[1]{Instituto Federal de Educa\c c\~ao, Ci\^encia e Tecnologia Baiano - Campus Senhor do Bonfim\\48970-000 Senhor do Bonfim, Bahia, Brazil}
\author[2]{Rossen Dandoloff}
\address[2]{Laboratoire de Physique Th\'eorique et Mod\'elisation, Universit\'e de Cergy-Pontoise\\
95302 Cergy-Pontoise, France}

\begin{abstract}We study the nonlinear $\sigma$-model in an external magnetic field applied on curved surfaces with rotational symmetry. The Euler-Lagrange equations derived from the Hamiltonian yield the double sine-Gordon equation (DSG) provided the magnetic field is tuned with the curvature of the surface. A $2\pi$ skyrmion appears like a solution for this model and surface deformations are predicted at the sector where the spins point in the opposite direction to the magnetic field.  We also study some specific examples by applying the model on three rotationally symmetric surfaces: the cylinder, the catenoid and the hyperboloid.\end{abstract}
\end{frontmatter}



The interplay between geometry and physical properties of condensed matter systems has demanded large attention in last decades. In part, this is associated to the growing capacity for fabricating and manipulating nanoscale devices with different geometries. For instance, recent developments in nanotechnology made it possible to produce quasi two-dimensional exotic shapes  like the M\"obius stripe \cite{Mobius-Nature}, torus \cite{Yazgan-Torus} and asymetric nanorings \cite{Zhu-PRL-96}. On the other hand, there is a growing theoretical interest in this subject due the fact that particle-like excitations, like vortices and skyrmions, for example, interact not only with each other, but also with the curvature of the substract \cite{Curv-Effect}. Vortices and skyrmions appear in several condensed matter systems, e.g. superconductors, BEC, nematic liquid crystals, ferro and antiferromagnets \cite{Chaikin-Book,Mermin-Review}. In the case of ferromagnetic materials, cylindrical nanomagnets with a vortex as the magnetization groundstate have been considered as candidates to be used in logic memory, data storage, highly sensitive sensors \cite{Dai-Appl} and cancer therapy \cite{Cancer-Therapy}. The energy of these topological excitations depends on the curvature of the ferromagnetic nanoparticle \cite{Vagson-JAP-105,Vagson-Submmited} and their dynamical properties are affected by the interaction with curve defects appearing during the fabrication of the nanomagnets \cite{Apolonio-JAP}.

In this context, an important issue is the control and manipulation of the shape and physical properties of nanoparticles. Indeed, several works have already devoted attention on this issue, e.g., theoretical works have shown that the application of a constant external magnetic field in a circular elastic cylinder surface coated by magnetic material must cause a deformation at the region where the spins are pointing in the opposite direction of the magnetic field \cite{Saxena-PRB-58}. At the same time, the concept of magnetoelastic metamaterials, which respond by being compressed by electromagnetic forces in the presence of an external magnetic or electric field, leads to a new generation of metamaterials that can be useful for the design of artificial media \cite{Lapine-Nature}. In addition, it was shown that the curvature of graphene bubbles can be controlled by applying an electric field \cite{georgiou-AIP} and still, by combining the curvature effects with magnetic fields, the molecular alignment of flux lines of the nematic director of nematic liquid crystals can be reoriented or switched between two stable configurations \cite{Napoli-PRL-108}.

In this letter, we show that the homogeneous DSG can be obtained for an arbitrary magnetically coated surface with rotational symmetry in the presence of an external magnetic field provided the field is tuned with the curvature of the surface. The obtained solution consists in a $2\pi$ skyrmion, which induces a deformation in the rotationally symmetric surfaces due the presence of the magnetic field. Besides, we apply the model to three specific cases:  cylinder, catenoid and hyperboloid, in order to obtain the characteristic lengths of the skyrmion for these particular surfaces. To our knowledge, this is the first time that the tuning between an external magnetic/electric field and the geometry of the substract yields a controllable system with well defined theoretical solutions. The coupling between the magnetic field and the geometry of condensed matter systems is a surprising result and we believe that this issue could guide future studies in the control and manipulation of the shape and physical properties of magnetoelastic coated surfaces, superfluid helium, nematic liquid crystals, graphene bubbles, topological insulators and so forth.

The total energy for a deformable, magnetoelastically coupled manifold is given by $E=E_{\text{mag}}+E_{\text{el}}+E_{\text{m-el}}$, where $E_{\text{mag}}$, $E_{\text{el}}$ and $E_{\text{m-el}}$ are the magnetic, elastic and magnetoelastic contributions to the total energy, respectively. The magnetic contribution is composed by the exchange, magnetostatic and Zeeman terms. In this work, we will focus our attention mainly on the exchange and Zeeman contributions to describe the magnetic properties of curved surfaces with rotational symmetry in the presence of an external magnetic field. In this case, the energy is well represented by the nonlinear $\sigma$-model on a surface in an external axial magnetic field:

\begin{equation}\label{GenHam}
H=\iint\left(\nabla\mathbf{m}\right)^2dS-g\mu\iint\mathbf{m}\cdot\mathbf{B}dS,
\end{equation}
where $\mathbf{m}$ is the magnetization unit vector, $dS$ is the surface element, $\mathbf{B}$ is the applied magnetic field, $\mu$ is the magnetic moment, and $g$ is the $g$ factor of the electrons in the magnetic material.

This model was previously considered for circular cylinder surface and the well known DSG \cite{Leung-PRB-26} was derived from the Euler-Lagrange equations \cite{Saxena-PRB-58}. In that case, the authors obtained a $2\pi$ skyrmion like solution and showed that a geometrical frustration appears due to a second length scale introduced by the magnetic field. On the other hand, unlike the cylinder case, when the above model was considered on an arbitrary curved surface, it yielded the inhomogeneous DSG, whose solution could be gotten only numerically \cite{Dandoloff-JPhysA}. Here, in order to obtain the homogeneous DSG system, we will consider that the magnetic field is in the $z$-axis direction and its strength is a function of $\rho$, that is, $\mathbf{B}\equiv B(\rho)\mathbf{z}$,  where $\rho\equiv\rho(z)$ is the radial distance from one point on the surface to the $z$ axis. 

To proceed with our analysis, it will be useful to rewrite Eq.(\ref{GenHam}) on a general geometry with metric tensor $g_{ij}$. In this case, the Hamiltonian is given by:
\begin{equation}\label{HamCurv}
H=\iint\sqrt{\frac{1}{g^{\rho\rho}g^{\phi\phi}}}\left[g^{\rho\rho}\left(\partial_\rho\Theta\right)^2+g^{\phi\phi}
\sin^2\Theta\left(\partial_\phi\Phi\right)^2+g\mu B(\rho)(1-\cos\Theta)\right]d\rho d\phi,
\end{equation}
where $g^{\rho\rho}$ and $g^{\phi\phi}$ are the contravariant metric elements. The magnetization unit vector is parametrized by $\mathbf{m}=(\sin\Theta\cos\Phi,\sin\Theta\sin\Phi,\cos\Theta)$ and the rotationally symmetric curved surfaces are parametrized in cylindrical coordinates system $(\rho,\phi,z)$.

The Euler-Lagrange equations for the Hamiltonian (\ref{HamCurv}) give us the following equations:
\begin{equation}\label{EulerTheta}
\sqrt{\frac{g^{\rho\rho}}{g^{\phi\phi}}}\partial_\rho\left(\sqrt{\frac{g^{\rho\rho}}{g^{\phi\phi}}}\partial_\rho\Theta\right)=\frac{\left(\partial_\phi\Phi\right)^2}{2}\sin2\Theta+\frac{1}{g^{\phi\phi}}B'(\rho)\sin\Theta
\end{equation}
and
\begin{equation}\label{EulerPhi}
\partial_\phi\left[\sqrt{\frac{g^{\phi\phi}}{g^{\rho\rho}}}\partial_\phi\Phi\right]=0,
\end{equation}
where $B'(\rho)\equiv g\mu B(\rho)$.

Once we are considering surfaces with rotational symmetry, the parametric equations associated to these can be given by $\mathbf{r}=(\rho\cos\phi,\rho\sin\phi,z(\rho))$,
where $\rho$ is the radius of the surface at height $z$, and $\phi$ accounts for the azimuthal angle. In this case, we have that the covariant metric elements are given by:
\begin{equation}
g_{\phi\phi}=\frac{1}{g^{\phi\phi}}=\rho^2\hspace{1cm}\text{and}\hspace{1cm} g_{\rho\rho}=\frac{1}{g^{\rho\rho}}=\left(\frac{\partial z}{\partial \rho}\right)^2+1,
\end{equation}
what implies that $\partial_\phi\left(\sqrt{\frac{g^{\phi\phi}}{g^{\rho\rho}}}\right)=0$. Thus, the Eq.(\ref{EulerPhi}) leads to:
\begin{equation}
\partial^2_\phi\Phi=0\rightarrow \Phi=\phi+\phi_0,
\end{equation}
where $\phi_0$ is a constant of integration that does not influences the energy calculations. Now, we can rewrite  Eq.(\ref{EulerTheta}) as:
\begin{equation}\label{GenThetaEuler}
\partial_\xi^2\Theta=\frac{\sin2\Theta}{2}+g_{\phi\phi}B'(\rho)\sin\Theta,
\end{equation}
where $d\xi=\sqrt{g^{\phi\phi}/g^{\rho\rho}}d\rho.$

From the Eq.(\ref{GenThetaEuler}), there are two situations to be considered: firstly, when we consider $B(\rho)=0$, it yields the sine-Gordon equation and a skyrmion, which satisfies the self-dual equations, is obtained \cite{cylinder}. This case was previously studied for many different curved geometries, e.g., cylinder \cite{cylinder,cylinder-2}, sphere \cite{sphere}, pseudosphere \cite{pseudosphere}, torus \cite{torus}, cone \cite{cone}, catenoid, hyperboloid \cite{Dandoloff-JPhysA,Vagson-Submmited}. It is important to note that, usually, the skyrmion presents a characteristic length scale (CLS) that depends on the length of the underlying manifold, which, in general, appears in front of the $\sin(2\Theta)$ term. However, in our model, this characteristic length is given by ($\sqrt{g^{\phi\phi}/g^{\rho\rho}}$), which is embedded in the $\xi$ parameter. Thus, the skyrmion would have its CLS rescheduled to one, that is, if we take the cylinder case, the CLS of the skyrmion is equal to the radius $\rho_{_0}$ of the cylinder. However, when working on an infinite cylinder, the change of variable $z\rightarrow z/\rho_{_0}$ eliminates any dependence on $\rho_{_0}$ \cite{cylinder}.

The second case to be analysed is given when we consider an arbitrary magnetic field $B(\rho)\neq0$. In this case, Eq.(\ref{GenThetaEuler}) gives the inhomogeneous DSG and numerical solutions are demanded \cite{Dandoloff-JPhysA}. However, note that if the external field is tuned with the surface curvature in the form $B'(\rho)=g^{\phi\phi}B'_0$, where $B'_0=g\mu B(\rho_{_0})$ and $\rho_{_0}$ is the surface radius at $z=0$ plane,  Eq.(\ref{GenThetaEuler}) yields the homogeneous DSG:
\begin{equation}\label{HomSGEq}
\partial_\xi^2\Theta=\frac{\sin2\Theta}{2}+B'_0\sin\Theta,
\end{equation}
whose solution can be given by:
\begin{equation}\label{solution}
\Theta(\xi)=2\tan^{-1}\left(\frac{\rho_{_B}}{\zeta\sinh\frac{\xi}{\zeta}}\right),
\end{equation}
where $\rho_{_B}^2=1/B'_0$ and $\zeta=\rho_{_B}/(1+\rho_{_B}^2)^{1/2}$. Then, one can see that the homogeneous DSG must be obtained provided that the magnetic field is tuned to the surface geometry, depending on the radius $\rho$.

The above equation represents a $2\pi$ skyrmion, which is a topological excitation belonging to the second class of the second homotopy group, whose CLS is given by $\zeta$. Note that the increasing of $B'_0$ leads to the decreasing of the CLS of the skyrmion causing it to remain confined in smaller regions of the surface. It is easy to note that $\zeta\rightarrow0$ when $B'_0\rightarrow\infty$. See also that if we do $B'_0=0$ in Eq.(\ref{solution}), we do not get the $\pi$ skyrmion, as it happens when we consider Eq.(\ref{HomSGEq}) with $B'_0=0$. 
This is explained because these two solutions belong to two different homotopy classes and cannot be transformed one to another by continuous transformation or by limiting process  $B'_0\rightarrow\infty$. By introducing a CLS in the system, the magnetic field induces a geometrical frustration, and the skyrmion, which had its CLS rescheduled to one, must choose a new CLS, given by $\zeta$, that is smaller than the length $\rho_{_B}$, introduced by the magnetic field.

In order to calculate the energy of the $2\pi$ skyrmion with the imposed conditions, we can rewrite the Hamiltonian (\ref{HamCurv}) as: 
\begin{equation}\label{GenHam2}
H'=H_1+H_2=2\pi\left[\int_{-\infty}^{\infty}\left[(\partial_\xi\Theta)^2+\sin^2\Theta\right]d\xi+
\int_{-\infty}^{\infty}\frac{1}{\rho_{_B}^2}(1-\cos\Theta)\right]d\xi,
\end{equation}
where $H_1$ is the part of the Hamiltonian that does not have dependence on the magnetic field and $H_2$ is the Zeeman term. In this case, the energy of the skyrmion is evaluated to give:
\begin{equation}\label{2piSolEn}
E_{_\text{S}}=8\pi\left[\left(1+\frac{1}{\rho_{_B}^2}\right)^{1/2}+\frac{1}{\rho_{_B}^2}\sinh^{-1}\rho_{_B}\right],
\end{equation}
that is larger than the minimum energy for the homotopy class with winding number $Q=2$, that is to say, $E_{_\text{2$\pi$ S}}=8\pi$. Only in the limit $\rho_{_B}\rightarrow\infty$, we get $E_{_\text{S}}\rightarrow8\pi$ and $\zeta\rightarrow1$. Once $\rho_{_B}=\sqrt{1/B'(\rho_{_0})}$, the energy of the skyrmion is associated with the strength of the magnetic field at the $z=0$ plane. In order to release the geometrical frustration introduced by the magnetic field, an elastic surface will deform to decrease the radius in the region where the skyrmion is centered. This will minimize the energy of the two terms that does not have dependence on the magnetic field in the Hamiltonian (\ref{GenHam2}).

Now, we will focus on the last term in the Hamiltonian $H'$, which involve the external magnetic field:
\begin{equation}
H_2=2\pi\int_{-\infty}^{\infty}\frac{1}{\rho_{_B}^2}(1-\cos\Theta) d\xi.
\end{equation}
For the solution (\ref{solution}), $H_2$ will take the form:
\begin{equation}
H_2=4\pi\int_{-\infty}^{\infty}\frac{1}{\rho_{_B}^2}\left[\frac{1}{1+\left(1-\zeta^2\right)\sinh^2\frac{\xi}{\zeta}}\right]d\xi=4\pi\int_{-\infty}^{\infty}f(\rho_{_B},\xi)d\xi.
\end{equation}
This part of the Hamiltonian will decrease in energy if we locally decrease $\xi$, but keep it constant in Eq.(\ref{solution}). Therefore, for a magnetic field tuned with the surface, if we keep the surface cross section at $\xi\pm\infty$, the surface will deform in the region of the skyrmion. In the case of small magnetic field, we get:
\begin{equation}
E_{_\text{S}}=8\pi\left[1+\frac{1}{2\rho_{_B}^2}\left(1+2\ln{2\rho_{_B}}\right)\right]
\end{equation}
and for large magnetic field, we get $E_{_\text{S}}=8\pi/\rho_{_B}$.

The highest contribution for the magnetic energy density due the interaction with the external magnetic field comes from the sector where the spins point in the opposite direction to the field. In this way, the $2\pi$ skyrmion would like to collapse and eliminate the region where the spins are opposite to the magnetic field. However, to align the spins, the system would fall in a $\pi$ skyrmion sector of the second homotopy group, which has infinite energy due the interaction of the spins at $+\infty$ with the magnetic field. Thus, there is a hard-core repulsion between the two $\pi$ skyrmions in the system by virtue of the curvature. Due the introduction of a geometric frustration in the problem by the magnetic field, the energy in Eq.(\ref{2piSolEn}) is grater than $8\pi$. Then, the surface will deform to decrease its energy till it have that one belonging to the second homotopy class \cite{Saxena-PRB-58}.

In the absence of an external field ($B_{_0}=0$), the self dual equation for the Eq.(\ref{HomSGEq}) is $\partial_\xi\Theta=\pm\sin\Theta$. For the DSG, we can write:
\begin{equation}\label{SDE}
\partial_{\xi}\Theta=\pm\sin\Theta\left[1+\frac{4\sin^2(\Theta/2)}{\rho_{_B}^2\sin^2\Theta}\right].
\end{equation}
The function in square brackets is greater than that one for all values of $\Theta(\xi)$ and the self dual equation is not satisfied. Therefore, the surface would deform in order to satisfy the self dual equation and to lower the energy. Note that all solutions of the Eq.(\ref{SDE}) satisfy the Euler-Lagrange equation, but not vice versa. The $2\pi$ skyrmion lattice of the double sine-Gordon equation in two regimes, $\rho_{_B}\leq1$ and $\rho_{_B}>1$, are similar to that found on the cylinder \cite{Saxena-PRB-58}.

Now, we will apply our generic model in some surfaces with rotational symmetry in order to find the skyrmion CLS associated to each one. Firstly, we will investigate the above model on the cylinder surface, which was previously studied by Saxena \textit{et al} \cite{Saxena-PRB-58}. In this case, we have that $z(\rho)\equiv z$ is a constant. This parametrization leads to $g_{\phi\phi}^\text{cyl}=\rho^2=r^2$ and $g_{\rho\rho}^\text{cyl}=1$, where $r$ is the radius of the cylinder. Thus the magnetic field to be applied on the surface would be $B'(\rho)_{_\text{cyl}}=({1}/{r^2})B'_0$. Since $r$ is constant, the external magnetic field is also a constant. Finally, we have that $\xi_{_\text{cyl}}=\ln{r}$. As expected, the result obtained here would be the same as that given in the Eq.(4) of Ref. \cite{Saxena-PRB-58} if we had considered $z$ instead $\rho$ as the variable of the problem. 

To our knowledge, unlike the cylinder, the homogeneous DSG for Heisenberg spins in an external field was not found in any other surface and attempts to find solutions to the model described by the equation (\ref{GenHam}) led to the inhomogeneous DSG, which must be solved numerically \cite{Dandoloff-JPhysA}. Here, we will focus our attention on two surfaces with negative varying Gaussian curvature, the catenoid and the hyperboloid. The catenoid is a minimal surface, as its mean curvature is zero everywhere. For this surface, we have that $\rho(z)=r\cosh(z/r)$, thus, this surface can be parametrized by $(\rho\cos\phi,\rho\sin\phi,r\cosh^{-1}(\rho/r))$, where $r$ is the radius of the catenoid at the $z=0$ plane. This parametrization leads to:
\begin{equation}
g_{\phi\phi}^\text{cat}=\rho^2\hspace{1cm}\text{and}\hspace{1cm} g_{\rho\rho}^\text{cat}=\frac{\rho^2}{\rho^2-r^2}.
\end{equation}
Then, the magnetic field to be applied on the surface, in order to obtain the homogeneous DSG, will be $B'(\rho)_{_\text{cat}}=({1}/{\rho^{2}})B'_0$. Once $\rho$ is not a constant for the catenoid, unlike the result obtained for the cylinder, the field must not be constant. Indeed, for $\rho\rightarrow\infty$, the field tends to zero and its maximum value ($B'(\rho=r)_{_\text{cat}}=({1}/{r^2})B'_0$) must occur at $\rho=r$. Finally, the catenoid in an external varying magnetic field admits a $2\pi$ skyrmion solution with \begin{equation}\xi_{_\text{cat}}=\ln\left[2\left(\rho+\sqrt{\rho^2-r^2}\right)\right].\end{equation}

The hyperboloid has a shape similar to the catenoid, however, while the catenoid has mean curvature null everywhere, the hyperboloid has variable mean and Gaussian curvatures. A hyperboloid can be parametrized by $(\rho\cos\phi,\rho\sin\phi,({b}/{r})\sqrt{\rho^2-r^2})$, where again $r$ is the radius of the surface at $z=0$ and $b$ is a multiplicative parameter that accounts for the height of the surface. This parametrization yields: 
\begin{equation}
g_{\phi\phi}^{\text{hyp}}=\rho^2\hspace{1cm}\text{and}\hspace{1cm} g_{\rho\rho}^{\text{hyp}}=\frac{\rho^2\left(b^2+r^2\right)-r^4}{r^2\left(\rho^2-r^2\right)}.
\end{equation}
From now on, we will use the biharmonic coordinate system (BC), in which $b=r$, to describe a particular kind of hyperboloid \cite{PolarHyp}, so called polar hyperboloid. In this case, the metric element is simplified to: 
 \begin{equation}
g_{\rho\rho}^{\text{phyp}}=\frac{2\rho^2-r^2}{\rho^2-r^2}
\end{equation}
and, to obtain the homogeneous DSG, the magnetic field must be given by $B'(\rho)_{_\text{hyp}}=({1}/{\rho^2})B'_0$. Thus, as well as the catenoid case, the field must be $\rho$ dependent, having its maximum value, given by $B'(\rho=r)_{_\text{hyp}}=({1}/{r^2})B'_0$, at the plane $z=0$. The magnetic field applied on the hyperboloid described by BC varies with $\rho=\sqrt{r^2+z^2}$. Then, $B'(\rho)_{_\text{hyp}}\rightarrow0$ when $z(\rho)\rightarrow\pm\infty$. Finally, the polar hyperboloid also admits the solution given in the equation (\ref{solution}), however, the $\xi$ parameter for this surface is given by a large and tedious expression in such a way that it will be omitted here. Despite the fact that $\lim_{\rho\rightarrow\infty}B(\rho)=0$, it is important to note that the total magnetic flux through these infinite surfaces diverges as $\ln\rho$ at infinity for the catenoid and hyperboloid, but remains constant and depending on $r$ for the cylinder. 

In conclusion, we have shown that the Euler-Lagrange equations derived from the continuum approximation for classical Heisenberg spins on a surface with rotational symmetry in an external magnetic field in the $z$ axis direction is the homogeneous DSG provided the field is tuned with the curvature of the surface. We have found a single $2\pi$ skyrmion-like solution to this model for an arbitrary surface and surface deformations at the sector where the spins point in the opposite direction of the magnetic field were predicted for an elastic surface. The model was applied to three specific cases: the cylinder, the catenoid and the hyperboloid, which have different characteristic length scales and curvatures. Each considered surface admits a skyrmion solution, but the magnetic field which gives the homogeneous DSG presents different behavior to each one, once it is a function of $\rho(z)$. While the field on the cylinder is constant, that one which leads to the DSG on the catenoid and the hyperboloid is inversely proportional to $\rho(z)$. Since the characteristic length of the skyrmion depends on the strength of the field, the magnitude of the deformation can be controlled by the magnitude of the applied magnetic field. 

These results may play an important role for experimental studies where the magnetoelastic effects predicted here could be observed. Besides, theoretical works including surface tension could clarify the deformation dynamics of magnetic coated membranes and/or magnetoelastic metamaterials \cite{Lapine-Nature}. Finally, we believe that this issue may be relevant for future studies on manipulation and control of the shape and physical properties of curved graphene sheets and topological insulators.

\section*{Acknowledgements}
We thank to the Brazilian agency CNPq (grant number 562867/2010-4) and Propes of the IF Baiano, for financial support. Carvalho-Santos thanks to F. A. Apolonio, J. D. Lima, P. G. Lima-Santos and G. H. Lima-Santos by encouraging the development of this work.

\end{document}